\documentstyle[12pt]{article}
\parskip=\medskipamount

\title{On consistency of the closed bosonic string with different
left-right ordering constants}
\author{A.A. Deriglazov\thanks{alexei@if.ufrj.br ~ On leave of
    absence from Dept. of Math. Phys., Tomsk Polytechnical University,
    Tomsk, Russia}}
\date{Instituto de F\'\i sica, Universidade Federal do Rio de Janeiro,\\
Rio de Janeiro, Brasil.}
\begin{document}
\maketitle
\large
\begin{abstract}
Closed bosonic string with different normal ordering constants  
$a \ne \bar a$ for the right and the left moving sectors is considered.
One immediate consequence of this choice is absence of tachyon in
the physical state spectrum. Selfconsistency of the resulting model in
the "old covariant quantization" (OCQ) framework is studyed. The model
is manifestly Poincare invariant, it has non trivial massless sector and 
is ghost free for $D=26, ~ a=1, ~\bar a=0$. A possibility to obtain  
the light-cone formulation for the model is also discussed. 
\end{abstract}

\noindent
{\bf PAC codes:} 11.25.Pm, 11.30.Pb  

For the case of the open bosonic string, presence of tachyon is necessary 
condition if one wishes to obtain a theory with a nontrivial massless 
sector presented. The same conclusion follows from a lot of
selfconsistency checks which can be performed by using of various 
quantization schemes [1-11]. For the closed string there is exist simple
possibility to remove tachyon from the physical spectrum. Namely, it
is sufficiently to choose different values of constants in the normal
ordered expressions for the constraints:
$L_0-a=0, ~ \bar L_0-\bar a=0, ~ a \ne \bar a$. Then the constraint 
$L_0-\bar L_0-(a-\bar a)=0$ prohibites appearance of the tachyon among the 
physical states. 

The aim of this Letter is to discuss selfconsistency of the resulting 
model. The model will be considered first in the OCQ framework. The massless 
sector is presented in the case of any nonnegative integer $a$, $\bar a$.
Further restrictions on the constants arise from the no ghost theorem. In
addition to the standard choise, one finds the unique possibility
 $a=1,~\bar a=0$. 
The ground state of the resulting theory turns out to be massless vector. 
Proof of the no ghost theorem will be discussed in some details (while the 
result $a=1, \bar a=0$ can be deduced, in principle, from the literature 
[1-6, 10-12]). It is motivated by the fact that the standard proof implies 
introduction of a sufficiently complicated machinery. In particular, one 
needs to consider the discretised (or the admissible [10]) momentum 
space and DDF states and then to establish relation  
among the DDF space and the total state space. Below more 
direct proof which do not involve of these tools will be presented.
Also, exact relation among the space-time dimension and the ordering
constants is find (see Eq.(\ref{100}) below).
In conclusion I consider also a possibility to remove anomaly in the
light cone Poincare algebra for the model under investigation. 
                          
In the gauge $g^{ab}=\eta^{ab}=(-,+)$ general solution of the 
equations of motion for the closed bosonic string is
(we use the standard conventions [10]: 
$x^\mu(\tau, \sigma+\pi)=x^\mu(\tau, \sigma), ~ 
\eta_{\mu\nu}=(-,+, \ldots ,+)$)
\begin{eqnarray}\label{2} 
x^\mu(\tau,\sigma)=X^\mu+\frac 1{\pi T}P^\mu\tau+\frac{i}{2\sqrt{\pi 
T}}\sum_{n\ne 0} 
\frac 1{n}\left[\bar\alpha^\mu_ne^{i2n(\tau+\sigma)}+
\alpha^\mu_{-n}e^{-i2n(\tau-\sigma)}\right]. 
\end{eqnarray}
Operators which correspond to the variables $(X,P,\alpha,\bar\alpha)$ will 
be denoted by the same symbols. Then the commutation relations are 
$[X^\mu,~P^\nu] \\
=-i\eta^{\mu\nu}, \quad 
[\alpha^\mu_n,~\alpha^\nu_k]=[\bar\alpha^\mu_n,~\bar\alpha^\nu_k]= - 
n\eta^{\mu\nu}\delta_{n+k,0}$,        
where $\alpha^\mu_0=-\bar\alpha^\mu_0\equiv\frac 1{2\sqrt{\pi T}}P^\mu$.   
General vector $\mid \Psi >$ of the total Fock space is linear combination of 
the basic vectors 
\begin{eqnarray}\label{4}
\mid \varphi >=\prod^{\infty}_{m,n=1}\prod^{D-1}_{\mu_m,\nu_n=0}
(\alpha_{-m}^{\mu_m})^{s_{m,\mu_m}}(\bar\alpha_{-n}^{\nu_n})^{t_{n,\nu_n}}
\mid 0, p >,
\end{eqnarray}                          
where $\mid 0, p >\equiv f(p)\mid 0 >$ and $P^\mu f(p)=p^\mu f(p)$. Beside  
that, one has the Virasoro operators which correspond to the first class 
constraints of the classical theory $(\partial_\tau x^\mu \pm 
\partial_\sigma x^\mu)^2=0$
\begin{eqnarray}\label{5}
L_m=\frac 12\sum_{\forall k}\alpha^\mu_{m-k}\alpha^\mu_k, \quad 
\bar L_m=\frac 12\sum_{\forall k}\bar\alpha^\mu_{m-k}\bar\alpha^\mu_k, \quad
m\ne 0,
\end{eqnarray}
\begin{eqnarray}\label{6}
L_0=\frac 1{8\pi T}P^2+N, \quad  
\bar L_0=\frac 1{8\pi T}P^2+\bar N,
\end{eqnarray}     
where the level number operators are
\begin{eqnarray}\label{7}
N=\sum_1^{\infty}\alpha^\mu_{-k}\alpha^\mu_k, \quad
\bar N=\sum_1^{\infty}\bar\alpha^\mu_{-k}\bar\alpha^\mu_k.
\end{eqnarray}   
Transition to the quantum theory is not a unique procedure since the 
quantities $L_0, \bar L_0$ involve products of the non-commuting operators.  
So one needs in some ordering prescription. An appropriate choice is normal 
ordering prescription (as it is written in Eq.(\ref{7})) which gives well 
defined operators for the Fock space realization. Then one more step is 
necessary in the OCQ framework. The normal ordering prescription implies   
appearance of anomaly terms in the quantum Virasoro algebra,
so one needs to use the Gupta-Bleuler quantization prescription by 
requiring that the physical states are annihilated by half of the operators  
(\ref{5}) only
\begin{eqnarray}\label{9}
L_m\mid ph >=\bar L_m\mid ph >=0, \quad  m>0,
\end{eqnarray}   
as well as by
\begin{eqnarray}\label{10}
(L_0-a)\mid ph >=(\bar L_0-\bar a)\mid ph >=0.
\end{eqnarray}     
Here the constants $a, \bar a$ (real numbers) correspond to the above 
mentioned ambiguouty and can be fixed further from the selfconsistency 
requirements. Since the theory is Poincare invariant, the Fock space
generated by
(\ref{4}) is a representation space of the Poincare group. In particular, 
the operator $P^\mu$ is identified with the Poincare shift generator. From 
this it follows that eugenvalue of the operator $-P^2$ in an irreducible 
subspace gives mass of the state: $M^2=-P^2$. Then Eq.(\ref{10}) can be 
rewritten in the form
\begin{eqnarray}\label{11}
M^2=4\pi T(N+\bar N-(a+\bar a)),
\end{eqnarray}  
\begin{eqnarray}\label{12}
N-\bar N-(a-\bar a)=0.
\end{eqnarray}   
The standard choice of the constants is, from the beginning: $a=\bar a$. 
Let me take them different. The immediate consequence is that the state \\ 
$\mid 0,p >\equiv f(p)\mid 0 >$ do not belong to the physical subspace
since it do not obeys to Eq.(\ref{12}). 

Our aim now is to discuss properties of this model. Consider conditions 
which can lead to appearance of the massless sector in the physical spectrum. 
Eugenvalues of the operators $N, \bar N$ in the space (\ref{4}) are integer 
non negative numbers, one writes $N\mid \varphi_0 >=(k+m)\mid \varphi_0 >, 
~ \bar N\mid \varphi_0 >= \\
m\mid \varphi_0 >$. From Eq.(\ref{12}) it follows
$a-\bar a=k$, if one
wishes to obtain non empty physical sector. From Eq.(\ref{11}) the mass of 
the state is $M^2=4\pi T\left(k+2m-(a+\bar a)\right)$. Thus the massless
sector can be presented if
\begin{eqnarray}\label{13}
a=k+m, \quad \bar a=m; \quad k\ge 0, \quad m\ge 0.
\end{eqnarray}       
An additional restrictions on $a, \bar a$ will follow from the no ghost
theorem.

It will be convenient to work in the light-cone coordinates: 
$\alpha_{-m}^\mu\equiv (\alpha_{-m}^+, \\
\alpha_{-m}^-, \alpha_{-m}^i)$. Since Eq.(\ref{4}) contains the commuting
operators only, they can be arranged in any
desired way. On the same reason it is sufficiently to consider only one 
(left or right) sector. Retaining only one sector, let me fix the following  
order (and notations) 
\begin{eqnarray}\label{14}
\mid \varphi_R >= 
\prod^\infty_{i=1}(\alpha^+_{-i})^{s_i}
\left[\prod^\infty_{k=1}\prod^{D-2}_{i_k=1}
(\alpha^{i_k}_{-k})^{\rho_{k,i_k}}\right]
(\alpha^-_{-1})^{\lambda_1}\ldots(\alpha^-_{-m})^{\lambda_m}\mid 0,~p > \cr
\equiv\mid (\alpha^-_{-1})^{\lambda_1},\ldots,
(\alpha^-_{-m})^{\lambda_m} >^\Lambda\equiv\mid ~ >^\Lambda,
\end{eqnarray}
where $\Lambda=\sum^m_{r=1}\lambda_r$ is a number of the operators
$\alpha^-$. Then the following statement can be formulated.\\
\underline{Lemma 1.} Any basic vector (\ref{14}) can be presented 
as a linear combination of the vectors 
\begin{eqnarray}\label{15}
\mid e_R >=L_{-1}^{\lambda_1}\ldots L^{\lambda_m}_{-m}
(\alpha^+_{-1})^{s_1}\ldots(\alpha^+_{-n})^{s_n}
\left[\prod^{\infty}_{k=1}\prod^{D-2}_{i_k=1}
(\alpha^{i_k}_{-k})^{\rho_{k,i_k}}\right]\mid 0,~p >\equiv \cr 
L^{\lambda_1}_{-1}\ldots L^{\lambda_m}_{-m} K^{s_1}_{-1}\ldots 
K^{s_n}_{-n}\mid t >.  
\end{eqnarray}
with some $(\lambda,s,\rho)$ (in general  case they are different from the 
corresponding factors in Eq.(\ref{14}). It was denoted  
$K_{-n}\equiv\alpha^+_{-n}$. Also, if
$L_0\mid \varphi_R >=R\mid \varphi_R >$, then
all the vectors $\mid e_R >$ are eigenvectors of $L_0$ with the same
eigenvalue
\begin{eqnarray}\label{16}
L_0\mid e_R >=R\mid e_R >, \quad \forall \quad\mid e_R > .
\end{eqnarray}

In other words {\it{all the operators}} $\alpha^-_{-r}$ can be
incorporated into $L_{-r}$, which is crucial property for establishing
of the no ghost theorem. The proof of Lemma 1 is as follows. Let me
start from the "higher order" operators $\alpha^-_{-m}$ which are
presented in Eq.(\ref{14}). One writes 
the identity\footnote{Appearance of the factor $\alpha^+_0$ in the 
denominator is the standard singularity of the light cone formulation [11].}
\begin{eqnarray}\label{17}
\alpha^-_{-m}\mid 0,p >=-\frac 1{\alpha^+_0}L_{-m}\mid 0,p >+
(\alpha^-_{-m}+\frac 1{\alpha^+_0}L_{-m})\mid 0,p >,
\end{eqnarray}
where $\alpha^+_0=\frac 1{2\sqrt{\pi T}}p^+\ne 0$. The bracket on the
r.h.s. do not contains $\alpha^-_{-m}$ (note that for $m>1$ it contains  
$\alpha^-_{-k}, ~ k<m$). Note  also that it contains in fact a finite 
number of terms only. By virtue of Eq.(\ref{17}) the basic vector 
(\ref{14}) can be presented as 
\begin{eqnarray}\label{18}
\mid ~ >^\Lambda=c\left[\prod\alpha^+\right]
\left[\prod\alpha^i\right](\alpha^-_{-1})^{\lambda_1}\ldots
(\alpha^-_{-m})^{\lambda_m-1}L_{-m}\mid 0,~p >+ \cr
\mid (\alpha^-_{-1})^{\lambda_1^{'}},\ldots ,
(\alpha^-_{-m+1})^{\lambda_{m-1}^{'}}, 
(\alpha^-_{-m})^{\lambda_m-1} >^\Lambda, 
\end{eqnarray}
where $c=-\frac 1{\alpha^+_0}$. Below we will omit unessential numerical 
factors arising in the process. Now the operator $L_{-m}$ can be moved
to the left. Since
$\alpha^\mu_{-n}L_{-m}=L_{-m}\alpha^\mu_{-n}-n\alpha^\mu_{-n-m}$, the
number $\Lambda$ in the process can not be changed, and the result is
of the form
\begin{eqnarray}\label{19}
\mid (\alpha^-_{-1})^{\lambda_1}, \ldots,
(\alpha^-_{-m})^{\lambda_m} >^\Lambda= \cr
L_{-m}\mid (\alpha^-_{-1})^{\lambda_1}\ldots
(\alpha^-_{-m})^{\lambda_m-1} >^{\Lambda-1}+
\mid ~ >^{\Lambda-1}+ \cr
\mid (\alpha^-_{-1})^{\lambda_1^{'}}, \ldots,
(\alpha^-_{-m+1})^{\lambda_{m-1}^{'}}, 
(\alpha^-_{-m})^{\lambda_m-1} >^\Lambda, 
\end{eqnarray}
where all the states $\mid ~ >$ on the right hand side are linear 
combinations of the vectors which have the form (\ref{14}). For the first 
two terms the number of the operators $\alpha^-$ is decreased by one unit.
(note that the second term on the r.h.s. contains in general
"higher order" operators
$\alpha^-_{-r},~r>m$). The last term has the same total number of the 
operators $\alpha^-$ as the initial vector, but {\it{ the number of the
"higher order" operators $\alpha^-_{-m}$ is decreased by one unit}}.
After numeruous
repetition of  the procedure for the last term one incorporates all the 
operators $\alpha^-_{-2}, \cdots ,\alpha^-_{-m}$ into 
$L_{-2},\cdots,L_{-m}$. Then (\ref{19}) acquires the form 
\begin{eqnarray}\label{20}
\mid ~ >^\Lambda=\sum_mL_{-m}\mid ~ >^{\Lambda-1}+\mid ~ >^{\Lambda-1}+
\mid (\alpha^-_{-1})^{\Lambda} >^\Lambda.
\end{eqnarray}
On the next step one notes that in the identity 
\begin{eqnarray}\label{21}
\alpha^-_{-1}=-\frac 1{\alpha^+_0}L_{-1}+
(\alpha^-_{-1}+\frac 1{\alpha^+_0}L_{-1}),
\end{eqnarray}
the bracket on the r.h.s. do not contains the operators $\alpha^-_{-m}$
at all. So on this stage number of the operators $\alpha^-$ in the last
term of Eq.(\ref{20}) begin to decrease, one finds 
\begin{eqnarray}\label{22}
\mid (\alpha^-_{-1})^\Lambda >^\Lambda=
L_{-1}\mid (\alpha^-_{-1})^{\Lambda-1} >^{\Lambda-1}+
\mid (\alpha^-_{-1})^{\Lambda-1} >^{\Lambda-1}=\ldots= \cr 
L_{-1}\mid (\alpha^-_{-1})^{\Lambda-1} >^{\Lambda-1}+
\mid ~ >^0=\ldots=L^\Lambda_{-1}\mid ~ >^0+\mid ~ >^0. 
\end{eqnarray}
Then Eq.(\ref{20}) acquires the form
\begin{eqnarray}\label{23}
\mid ~ >^\Lambda=\sum_mL_{-m}\mid ~ >^{\Lambda-1}+\mid ~ >^{\Lambda-1}+
\mid ~ >^0,
\end{eqnarray}
where all the vectors of the form $L^\Lambda_{-1}\mid ~ >^0$ were included
into $\sum L_{-m}\mid ~ >^{\Lambda-1}$. In the result all the vectors
$| >$ on the r.h.s. of Eq.(\ref{23}) are of the form (\ref{4}), but
contain less then $\Lambda$ operators.

Now all the previous procedure can be numeruosly repeated for the vectors 
$\mid ~ >^{\Lambda-1}$ from (\ref{23}), with the final result being of the
form (\ref{15}). One notices also that the total level number $R$ can not be 
changed in the process described, so Eq.(\ref{16}) holds.

Let me introduce subspaces of (\ref{15}) as follows 
\begin{eqnarray}\label{24}
K_R=\left\{\mid k >=K^{s_1}_{-1}\ldots K^{s_n}_{-n}\mid t >\right\},
\end{eqnarray}
\begin{eqnarray}\label{25}
S_R=\left\{\mid s >=L^{\lambda_1}_{-1}\ldots L^{\lambda_m}_{-m}
K^{s_1}_{-1}\ldots K^{s_n}_{-n}\mid t >, \quad
\sum^m_{r=1}r\lambda_r>0\right\}.
\end{eqnarray}
By construction $S_R$ consist of the spurious vectors.
Evidently, the space (\ref{15}) is a direct sum of (\ref{24}),(\ref{25}).
Further, the transverse vectors $|t>$ have non negative norm and obey
the property $K_n|t>=0,\quad n>0,$ from which it follows \\        
\underline{Lemma 2.} Vectors of the subspace (\ref{24}) have
nonnegative norm. In particular, $< k\mid k >=0$ if $\sum^n_{r=1}rs_r>0$.\\  
Below we use also the following result.\\                                 
\underline{Lemma 3.} Vectors of the subspace (\ref{24}) are linearly
independent from the vectors of (\ref{25}).\\                         
Actually, suppose an opposite 
\begin{eqnarray}\label{26}
\mid s >=L^{\lambda_1}_{-1}\ldots L^{\lambda_m}_{-m}
K^{s_1}_{-1}\ldots K^{s_n}_{-n}\mid t >=
\sum c(\{l\},t^{'})K^{l_1}_{-1}\cdots K^{l_k}_{-k}\mid t^{'} >.
\end{eqnarray}
where $\sum^m_{r=1}r\lambda_r\equiv s>0$ and consider action of the operator 
$K_s$ on both sides of (\ref{26}). Then the r.h.s. is zero, while the l.h.s. 
can be computed directly, one finds  
\begin{eqnarray}\label{27}
K_s\mid s >\equiv c\frac{p^+}{2\sqrt{\pi T}} K^{s_1}_{-1}\ldots
K^{s_n}_{-n}\mid t >=0,
\end{eqnarray}
where $c>0$. From this it follows $\mid s >=0$. \\
These results allows one to prove the no ghost theorem.\\
\underline{Theorem.} Vectors of the physical subspace 
(\ref{9}), (\ref{10}) have nonnegative norm for the case
\begin{eqnarray}\label{100}
a\le 1, \quad  \bar a\le 1, \quad
D\le\frac{2(2-a)(21-8a)}{3-2a}, \quad
D\le\frac{2(2-\bar a)(21-8\bar a)}{3-2\bar a}.
\end{eqnarray}

The proof is as follows. The physical state can be presented as combination 
of the basic vectors (\ref{4}): $\mid ph >=\sum c_i\mid \varphi_i >$.
By using of linear independence of the vectors $\mid \varphi_i >$ one
concludes that all of them
obey to Eq.(\ref{10}) with the same eugenvalue as the vector
$\mid ph >$. In its turn, $\mid \varphi_i >$ can be presented as combination
of the vectors $\mid e_R >\times\mid e_L >$
according to Lemma 1. All of them obey also to Eq.(\ref{10}). Thus one 
writes $\mid ph >=\mid k >+\mid s >$,
$\mid k >\subset K\equiv K_R\times K_L$,
$\mid s >\subset S\equiv S_R\times S_L$, where $\mid k >,~ \mid s >$ obey
to
Eq.(\ref{10}). Further, the operator $L_{-m}(\bar L_{-m})$ with $m>2$ can 
be presented through $L_{-1}, L_{-2}, (\bar L_{-1}, \bar L_{-2})$ or, 
equivalently, through
\footnote{Application of the operators $\tilde L_{\pm2}$ instead of
$L_{\pm2}$ is the standard technical moment. The operator $\tilde L_2$
is chosen in such a way that
$[\tilde L_2,~L_{-1}]\mid \chi_1 >=0$, which simplifies the
proof below.} $L_{-1}, \tilde L_{-2}$ $(\bar L_{-1}, \tilde{\bar L}_{-2})$, 
where  
\begin{eqnarray}\label{29}
\tilde L_{\pm 2}=L_{\pm 2}+cL^2_{\pm 1}, \quad
c=\frac 3{2(3-2a)}; \cr
\tilde{\bar L}_{\pm 2}=\bar L_{\pm 2}+\bar c\bar L^2_{\pm 1}, \quad
\bar c=\frac 3{2(3-2\bar a)}.
\end{eqnarray}        
It allows one to rewrite the physical state as 
\begin{eqnarray}\label{30}
\mid ph >=\mid k >+L_{-1}\mid \chi_1 >+
\tilde L_{-2}\mid \chi_2 >+\bar L_{-1}\mid \bar\chi_1 >+
\tilde{\bar L}_{-2}\mid \bar\chi_2 >.
\end{eqnarray}        
The norm of the state is
\begin{eqnarray}\label{31}
< ph \mid ph >=< k \mid k >+< \chi_1 \mid L_1|k >+
< \chi_2 \mid \tilde L_2 \mid k >+ \cr
< \bar\chi_1 \mid \bar L_1| k>+
< \bar\chi_2 \mid \tilde{\bar L_2} \mid k >.
\end{eqnarray}        
Thus one needs to know $L_i|k>$. This information can be obtained
from the condition that $|ph>$ is the physical vector. Namely, one
finds after some algebra
\begin{eqnarray}\label{32}
L_1\mid ph >=0 ~ \Longrightarrow ~
\mid k_1 >-2(1-a)\mid \chi_1 >+\mid s_1 >=0,
\end{eqnarray}        
\begin{eqnarray}\label{33}
\tilde L_2\mid ph >=0 ~ \Longrightarrow ~
\mid k_2 >-A\mid \chi_2 >+\mid s_2 >=0,
\end{eqnarray}        
\begin{eqnarray}\label{34}
\bar L_1\mid ph >=0 ~ \Longrightarrow ~ \mid \bar k_1 >-
2(1-\bar a)\mid \bar\chi_1 >+\mid \bar s_1 >=0,
\end{eqnarray}        
\begin{eqnarray}\label{35}
\tilde{\bar L}_2\mid ph >=0 ~ \Longrightarrow ~ \mid \bar k_2 >-
\bar A\mid \bar\chi_2 >+\mid \bar s_2 >=0,
\end{eqnarray}        
where
\begin{eqnarray}\label{36}
\mid k_1 >=L_1\mid k >, \quad \mid k_2 >=\tilde L_2\mid k >, \quad
\mid k_i >\subset K; \cr
\mid \bar k_1 >=\bar L_1\mid k >, \quad \mid \bar k_2 >=
\tilde{\bar L}_2\mid k >, \quad
\mid \bar k_i >\subset K; \cr
\mid s_i >\subset S, \quad
\mid \bar s_i >\subset S. 
\end{eqnarray}   
It was also denoted
\begin{eqnarray}\label{37}
A(c, a)=4(c^2+3c+1)(2-a)-8c^2(2-a)^2-\frac 12 D, \quad
\bar A\equiv A(\bar c, \bar a). 
\end{eqnarray}        
It remains to consider various possibilities for the vectors
$\mid \chi_i >, \mid \bar\chi_i >$. Let $\mid \chi_1 >$ do not contains
of the operators
$L_{-m}, \bar L_{-n}:~ \mid \chi_1 >=\mid k_0 >\subset K$,
while $\mid \chi_2 >, \mid \bar\chi_1 >, \mid \bar\chi_2 >\subset S$.
From Lemma 3 and Eqs.(\ref{32})-(\ref{36}) it follows 
\begin{eqnarray}\label{38}
L_1\mid k >=2(1-a)\mid k_0 >, \quad \mid s_1 >=0, \cr
\tilde L_2\mid k >=\bar L_1\mid k >=\tilde{\bar L_2}\mid k >=0, \cr
\tilde L_2\mid s >=\bar L_1\mid s >=\tilde{\bar L_2}\mid s >=0. 
\end{eqnarray}
While $\mid s >$ is not a physical state, Eq.(\ref{38}) allows one
to estimate the norm (\ref{31}), namely
\begin{eqnarray}\label{39}
< ph \mid ph >=< k \mid k >+2(1-a)< k_0 \mid k_0 >.
\end{eqnarray}        
According to Lemma 2, the condition $a\le 1$ garantees that
$< ph \mid ph > \ge 0$.

Other possibilities can be analysed in a similar way. The result is that the 
conditions $1-a\ge 0$, $1-\bar a\ge 0$, $A\ge 0$, $\bar A\ge 0$ 
garantee non negative norm of the physical state. From this it follows 
Eq.(\ref{100}). 

Let me return to the discussion of the model (\ref{2}),(\ref{13}).
Besides the standard choice $a=\bar a=1$, the only other possibility
which is consistent with the no-ghost theorem is \footnote{An equivalent
choice is, of course, $a=0$, $\bar a=1$.} $a=1$, $\bar a=0$, which
implies also $D\le 26$ (see Eqs.(\ref{13}), (\ref{100})). The
requirement of absence of the Weyl anomaly implies $D=26$.
Now the lowest mass sector is the massless one and is selected by
\begin{eqnarray}\label{41}
(N-\bar N-1) \mid \varphi_0 >=0, \quad  (N+\bar N-1)\mid \varphi_0 >=0,
\end{eqnarray}        
as well as by Eq.(\ref{9}). The standard reasoning [12] shows that the
ground state is the equivalence class
\begin{eqnarray}\label{42}
\widetilde{\mid \varphi_0 >}=\left\{[f_\mu(p)+
d(p)p_\mu]\alpha^\mu_{-1}\mid 0 >, \quad p^2=0, \quad 
f_\mu p^\mu=0, \quad \forall d(p)\right\},
\end{eqnarray}        
which corresponds to the massless vector particle with $D-2$ transverse
physical polarisations. In the result the closed string with $a=1$,
$\bar a=0$ has the same massless sector as the open string. The second,
third, $\ldots$ massive levels consist of states of the form
$(\alpha_{-1})^2\bar\alpha_{-1}\mid 0 >$,
$(\alpha_{-1})^3(\bar\alpha_{-1})^2\mid 0 >$, $\ldots ~ $.

Thus, in the OCQ framework the closed bosonic string for the case
$a=1$, $\bar a=0$, $D=26$ is the ghost free manifestly Poincare invariant
theory without tachyon and with massless vector particle being its
ground state. Note also that it seems to be consistent at least on the
level of tree amplitudes since the corresponding calculations can be
performed in the OCQ framework [9].

In conclusion, consider a possibility of the light cone formulation for the 
model. In this case one impose the gauge $\alpha^+_m=\bar\alpha^+_m=0$, 
$m\ne 0$, $X^+=0$ for the constraints $L_m=\bar L_m=L_0+\bar L_0=0$ of the 
classical theory. Then the remaining physical degrees of freedom are $X^-, 
P^+, X^i, P^i, \alpha^i_m, 
\bar\alpha^i_m$. The Dirac bracket for these variables coincides with
the Poisson one. The remaining constraint involve the transverse
oscillators only [11]
\begin{eqnarray}\label{43}
L_{0, tr}-\bar L_{0, tr}-\bar b=0.
\end{eqnarray}        
Here some ordering constant $\bar b$ was included. In this formulation
the ghosts are absent by construction, but one needs to check Poincare
invariance of the resulting quantum theory. The only dangerous
commutator of the Poincare algebra is known to be $[J^{i-}, J^{j-}]$,
which must be zero. Manifest form of the generators are
\begin{eqnarray}\label{44}
J^{i-}=\frac{2\pi T}{p^+}(L_{0, tr}+\bar L_{0, tr}-b)-X^-P^i+
\frac{2i\sqrt{\pi T}}{p^+}(S^{i-}-\bar S^{i-}),
\end{eqnarray}        
where $S^{i-}=\sum_{n=1}^\infty\frac 1n(\alpha_{-n}^iL_{n, tr}-
L_{-n, tr}\alpha^i_n)$, and $b$ is the second ordering constant 
of the light cone formulation.
For the case $D=26$ the commutator can be presented as
\begin{eqnarray}\label{45}
[J^{i-},~J^{j-}]=\frac{4\pi T}{(p^+)^2}
\left\{(L_{0, tr}-\bar L_{0, tr}+b)S^{ij}-
(L_{0, tr}-\bar L_{0, tr}-b)\bar S^{ij}-\right. \cr
\left. 2S^{ij}-2\bar S^{ij}\right\},
\end{eqnarray}        
where the last two terms is the anomaly resulting from reordering of
operators
in the process. The standard choice is $b=2$, $\bar b=0$, which gives the 
desired result $[J^{i-}, J^{j-}]=0$ as a consequence of Eq.(\ref{43}). Other 
formal possibility is as follows: consider the normal ordering prescription 
for the operators $\alpha^i_m$ and the Weyl prescription for 
$\bar\alpha^i_m$. Then the anomaly terms in the left sector are absent, and 
the commutator is proportional to $(L^N_{0, tr}-\bar 
L^W_{0, tr}+b)S^{ij}-(L^N_{0, tr}-\bar L^W_{0, tr}-b)\bar S^{ij}- 2S^{ij}$.
It will be zero if one takes $b=\bar b=1$. 

It is known that the Weyl quantization (of the both sectors) for the
ordinary string is not consistent (this fact can be extracted also from
Eq.(\ref{45})). The problem of the Weyl quantization was investigated
also in context of the null string [13-16]. From the
previous discussion it follows that namely the Weyl quantization can be 
interesting for the light cone formulation of the model under investigation. 
This problem will be considered in a separate publication.

\section*{Acknowledgments.}

Author thanks N. Berkovits for useful discussions.
The work has been supported by FAPERJ and partially by
Project INTAS-96-0308 and by Project GRACENAS 
No 97-6.2-34.

\end{document}